\documentclass[pra,preprint,superscriptaddress,floatfix,showpacs]{revtex4-1}

\usepackage{amsmath,amsfonts,amssymb}
\usepackage{graphicx}

\def\beq{\begin{equation}}
\def\eeq{\end{equation}}
\def\beqn{\begin{eqnarray}}
\def\eeqn{\end{eqnarray}}

\def\bl {\mbox{\boldmath $[$}}
\def\br {\mbox{\boldmath $]$}}
\def\r {{\bf r}}
\def\p {{\bf p}}

\begin{document}

\title{Uncertainty product of an out-of-equilibrium\\ Bose-Einstein condensate}
\author{Shachar Klaiman}
\affiliation{Theoretische Chemie, Physikalisch--Chemisches Institut, Universit\"at Heidelberg, 
Im Neuenheimer Feld 229, D-69120 Heidelberg, Germany}
\author{Alexej I. Streltsov}
\affiliation{Theoretische Chemie, Physikalisch--Chemisches Institut, Universit\"at Heidelberg, 
Im Neuenheimer Feld 229, D-69120 Heidelberg, Germany}
\author{Ofir E. Alon}
\email{ofir@research.haifa.ac.il}
\affiliation{Department of Physics, University of Haifa at Oranim, Tivon 36006, Israel}
\date{\today}

\begin{abstract}
The variance and uncertainty product of
the position and momentum many-particle operators
of structureless bosons
interacting by a long-range inter-particle interaction 
and trapped in a single-well potential are investigated.
In the first example, of an out-of-equilibrium interaction-quench scenario,
it is found that, despite the system being fully condensed, 
already when a fraction of a particle is depleted differences
with respect to the mean-field quantities emerge.
In the second example, 
of the pathway from condensation to fragmentation of the ground state,
we find out that, although the cloud's density broadens while the system's fragments,
the position variance actually decreases, the momentum variance increases,
and the uncertainty product is not a monotonous function but has a maximum.
Implication are briefly discussed.
\end{abstract}

\pacs{03.75.Kk, 67.85.De, 03.75.Hh, 67.85.Bc, 03.65.-w}

\maketitle 

\section{Introduction}\label{Intro}

Bose-Einstein condensates (BECs) made of ultracold trapped 
bosonic atoms have become a vast ground to study interacting 
quantum systems \cite{ex1,ex2,ex3,NB1,NB2,rev1,rev2,rev3,rev4,book1,book2,book3}.
There has been a substantial theoretical interest in BECs,
and ample studies have been made 
to describe their static and dynamic properties using Gross-Pitaevskii theory. 
The time-dependent Gross-Pitaevskii equation governs a mean-field theory which
assumes that all boson are described by a single time-dependent one-particle function
throughout the evolution of the BEC in time.
It is generally accepted that Gross-Pitaevskii theory adequately describes
the ground state and out-of-equilibrium dynamics of BECs in the limit of an infinite number of particles
and at constant interaction parameter
(i.e., when the product of the number of particles times the scattering length is kept fixed).
Here, mathematically rigorous results exist and show that, under certain conditions,
the energy per particle and density per particle of the many-boson system coincide in this limit
with the Gross-Piteavekii results 
and that the system is $100\%$ condensed \cite{Yngvason_PRA,Lieb_PRL,Erdos_PRL,MATH_ERDOS_REF}.

Whereas a condensate fraction of 100\% 
implies that the number of depleted (non-condensed) particles divided by the total number of particles 
vanishes in the infinite-particle limit,
the former is always non-zero in an interacting many-boson system.
This observation has motivated us recently to look at properties of BECs 
which depend on the number of depleted particles rather than the condensate fraction.  
In \cite{Variance, TD_Variance} we showed, for the ground state as well as for an out-of-equilibrium BEC,
that even in the infinite-particle limit when the BEC is 100\% condensed, 
the variance of a many-particle operator and the uncertainty product of
two such operators can differ from those predicted by the Gross-Pitaevskii theory. 
The existence of many-body effects beyond those predicted by Gross-Pitaevskii theory stems 
from the necessity of performing the infinite-particle limit only after a many-particle 
quantum mechanical observable is evaluated and not prior to its evaluation.
Unlike the variance of operators of a single particle \cite{QM_book},
the variance and uncertainty product of many-particle operators
is more involved, 
also see \cite{Drumm,Oriol_Robin,BB_HIM} in this context.
It has furthermore been shown that the overlap of the exact and Gross-Pitaevskii wave-functions
of a trapped BEC in the limit of an infinite number of particles 
is always smaller than unity and may even become vanishingly small \cite{SL_Psi}.

The purpose of the present work is to build on and go beyond \cite{Variance,TD_Variance}
in two directions, first, by studying the variance and uncertainty product of
trapped bosons with a long-range inter-particle interaction and, second, 
when the system is no longer condensed.
The structure of the paper is as follows.
In Sec.~\ref{Theory} we briefly 
discuss a general theory
for the many-body variance and uncertainty product of an out-of-equilibrium trapped BEC.
In Sec.~\ref{APPL} we present two applications,
for the breathing dynamics of a trapped BEC (Subsec.~\ref{Ex1})
and for the pathway from condensation to fragmentation 
of the ground state of trapped interacting bosons (Subsec.~\ref{Ex2}). 
Concluding remarks are put forward in Sec.~\ref{Conclusions}.
Numerical and convergence details are collected in the Appendix.

\section{Theoretical Framework}\label{Theory}

Consider the many-body Hamiltonian of $N$ interacting bosons in a trap $V(\r)$,
\beq\label{HAM}
 \hat H(\r_1,\ldots,\r_N) = 
\sum_{j=1}^N \left[-\frac{1}{2} \frac{\partial^2}{\partial \r_j^2} 
+ \hat V(\r_j)\right] + \sum_{j<k} \lambda_0\hat W(\r_j-\r_k).
\eeq
Here, $\hbar=m=1$,
and $\hat W(\r_1-\r_2)$ is the inter-particle interaction with $\lambda_0$ its strength.

The system evolves in time according to the 
time-dependent Schr\"odinger equation,
\beq\label{TDSE}
\hat H(\r_1,\ldots,\r_N) \Psi(\r_1,\ldots,\r_N;t) = i \frac{\partial\Psi(\r_1,\ldots,\r_N;t)}{\partial t},
\eeq
where the wave-function $\Psi(\r_1,\ldots,\r_N;t)$ is normalized to unity.
Typically, the system is initially prepared in the ground state of the trap $V(\r)$,
$\hat H(\r_1,\ldots,\r_N) \Phi(\r_1,\ldots,\r_N) = E \Phi(\r_1,\ldots,\r_N)$,
where $E$ is the ground-state energy and $\Phi(\r_1,\ldots,\r_N)$ normalized to unity.

In what follows we employ the reduced one-body and two-body 
density matrices of $\Psi(\r_1,\ldots,\r_N;t)$ \cite{Lowdin,Yukalov,Mazz,RDMs}.
The reduced one-body density matrix is given by
\beqn\label{1RDM}
\frac{\rho^{(1)}(\r_1,\r_1';t)}{N} &=&
\int d\r_2 \ldots d\r_N \, \Psi^\ast(\r_1',\r_2,\ldots,\r_N;t) \Psi(\r_1,\r_2,\ldots,\r_N;t) = \nonumber \\
 &=& \sum_j \frac{n_j(t)}{N} \, \alpha_j(\r_1;t) \alpha^\ast_j(\r'_1;t).
\eeqn
The quantities $\alpha_j(\r;t)$ are the so-called natural orbitals and $n_j(t)$ their respective occupations
which are time dependent and used to define the
degree of condensation 
in a system of interacting bosons \cite{Penrose_Onsager}.
The density of the system is simply the diagonal of the reduced one-body density matrix,
$\rho(\r;t) = \rho^{(1)}(\r,\r;t)$.

We express in what follows quantities using the time-dependent natural orbitals $\alpha_j(\r;t)$.
The diagonal part of the reduced two-body density matrix is given by
\beqn\label{2RDM}
\frac{\rho^{(2)}(\r_1,\r_2,\r_1,\r_2;t)}{N(N-1)} &=& 
 \int d\r_3 \ldots d\r_N \, \Psi^\ast(\r_1,\r_2,\ldots,\r_N;t) \Psi(\r_1,\r_2,\ldots,\r_N;t) = \nonumber \\
 &=& \sum_{jpkq} \frac{\rho_{jpkq}(t)}{N(N-1)} \, 
\alpha^\ast_j(\r_1;t) \alpha^\ast_p(\r_2;t) \alpha_k(\r_1;t) \alpha_q(\r_2;t), 
\eeqn
where the matrix elements are 
$\rho_{jpkq}(t) = \langle\Psi(t)|\hat b_j^\dag \hat b_p^\dag \hat b_k \hat b_q|\Psi(t)\rangle$,
and the creation and annihilation operators are associated with the time-dependent natural orbitals $\alpha_j(\r;t)$.

To express the variance and uncertainty product of operators 
we begin simply with the operator
\beq\label{lin_term}
 \hat A = \sum_{j=1}^N \hat a(\r_j)
\eeq
of the many-particle system,
where $\hat a(\r)$ is a Hermitian operator.
A straightforward calculation gives 
the average per particle of $\hat A$ in the state $|\Psi(t)\rangle$,
\beq\label{Average_A}
\frac{1}{N}\langle\Psi(t)|\hat A|\Psi(t)\rangle = \int d\r \frac{\rho(\r;t)}{N} a(\r),
\eeq
which is directly expressed in terms of the system's density per particle
in case of a local operator in coordinate space $\r$.
In the case of a local operator in momentum space $\p$
we can write 
$\frac{1}{N}\langle\Psi(t)|\hat A|\Psi(t)\rangle = \int d\p \frac{\rho(\p;t)}{N} a(\p)$.

To proceed we also need the expectation value of the square of $\hat A$, 
\beq\label{sqr_term}
 \hat A^2 = \sum_{j=1}^N \hat a^2(\r_j) + \sum_{j<k} 2 \hat a(\r_j) \hat a(\r_k),
\eeq
which is comprised of one-body {\it and} two-body operators.
Expressed in terms of the density per particle and natural orbitals, 
the time-dependent variance per particle of the operator $\hat A$
can be written as a sum of two terms
\beqn\label{dis}
& & \frac{1}{N}\Delta_{\hat A}^2(t) = \frac{1}{N} 
\left[\langle\Psi(t)|\hat A^2|\Psi(t)\rangle - \langle\Psi(t)|\hat A|\Psi(t)\rangle^2\right] \equiv 
\Delta_{\hat a, density}^2(t) + \Delta_{\hat a, MB}^2(t), \nonumber \\
& & \quad \Delta_{\hat a, density}^2(t) = 
\int d\r \frac{\rho(\r;t)}{N} a^2(\r) - \left[\int d\r \frac{\rho(\r;t)}{N} a(\r) \right]^2, \nonumber \\ 
& & \quad \Delta_{\hat a, MB}^2(t) = \frac{\rho_{1111}(t)}{N} \left[\int d\r |\alpha_1(\r;t)|^2 a(\r) \right]^2 
- (N-1) \left[\int d\r \frac{\rho(\r;t)}{N} a(\r) \right]^2 + \nonumber \\
& & \quad \quad + \sum_{jpkq\ne 1111} \frac{\rho_{jpkq}(t)}{N} \left[\int d\r \alpha^\ast_j(\r;t) \alpha_k(\r;t) a(\r) \right]
\left[\int d\r \alpha^\ast_p(\r;t) \alpha_q(\r;t) a(\r)\right]. \
\eeqn
The first term, which is denoted by $\Delta_{\hat a, density}^2(t)$,
describes the variance of $\hat a(\r)$ resulting {\it solely} from the density per particle $\frac{\rho(\r;t)}{N}$.
The second term, which is denoted by $\Delta_{\hat a, MB}^2(t)$,
collects all other contributions to the many-particle variance and
is identically zero within Gross-Pitaevskii theory.
$\Delta_{\hat a, MB}^2(t)$ is generally non-zero within a many-body theory.

Finally, taking two many-particle operators,
\beq
\hat A=\sum_{j=1}^N \hat a(\r_j), \qquad \hat B=\sum_{j=1}^N \hat b(\r_j),
\eeq
and their respective time-dependent variances per particle, 
$\frac{1}{N}\Delta_{\hat A}^2(t)$ and $\frac{1}{N}\Delta_{\hat B}^2(t)$,
their uncertainty product satisfies the inequality 
\beq\label{uncertainty_t_1}
\frac{1}{N}\Delta_{\hat A}^2(t) \frac{1}{N}\Delta_{\hat B}^2(t) \equiv
\Delta_{\frac{\hat A}{N}}^2(t) \Delta_{\hat B}^2(t)
\ge \frac{1}{4}\left|\int d\r \frac{\rho(\r;t)}{N} \bl\hat a(\r),\hat b(\r)\br\right|^2,
\eeq
where $\bl\hat a(\r),\hat b(\r)\br$ is the commutator of the 
two Hermitian operators $\hat a(\r)$ and $\hat b(\r)$.

\section{Applications}\label{APPL}

As mentioned above we would like to go beyond \cite{Variance,TD_Variance}
in the investigation of the variance and uncertainty product of trapped BECs.
To this end, we here study trapped bosons with a long-range interaction 
and when the system is no longer condensed.
We treat structureless bosons with harmonic inter-particle interaction trapped
in a single-well anharmonic potential. 
The time-dependent Schr\"odinger equation of the trapped BEC
has no analytical solution in the present study, see in this respect \cite{SL_Psi,Cohen},
nor even the variance and uncertainty product can be computed analytically,
thus a numerical solution of the out-of-equilibrium dynamics is a must.
This will lead to interesting results.

We need a suitable and proved many-body tool to make the calculations.
Such a many-body tool is the multiconfigurational time-dependent
Hartree for bosons (MCTDHB) method,
which has been well documented \cite{MCTDHB1,MCTDHB2,book_MCTDH,Kaspar_The,book_nick,Axel_The},
benchmarked \cite{Benchmarks}, 
and extensively used 
\cite{BJJ,MCTDHB_OCT,MCTDHB_Shapiro,LC_NJP,Breaking,jcs1,Peter_2015a,Peter_2015b,Uwe,Sven_Tom,
Tunneling_Rapha,Peter_b,Alexej_u,jcs2,Kaspar_n,jcs3,MCTDHB_spin,jcs4,Axel_ar} 
in the literature.

\subsection{Breathing dynamics}\label{Ex1}

\begin{figure}[!]
\hglue -1.0 truecm
\includegraphics[width=0.345\columnwidth,angle=-90]{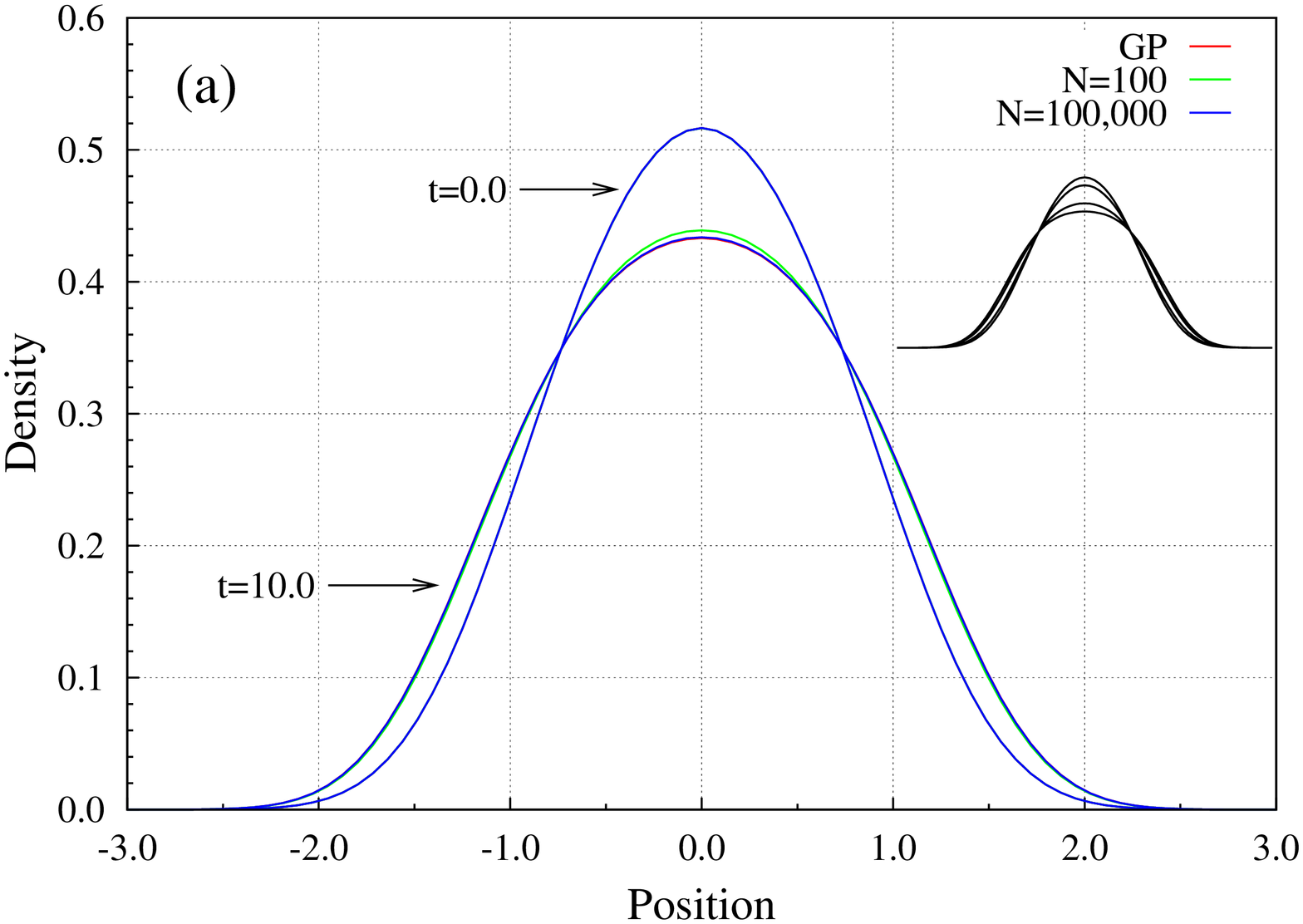}
\includegraphics[width=0.345\columnwidth,angle=-90]{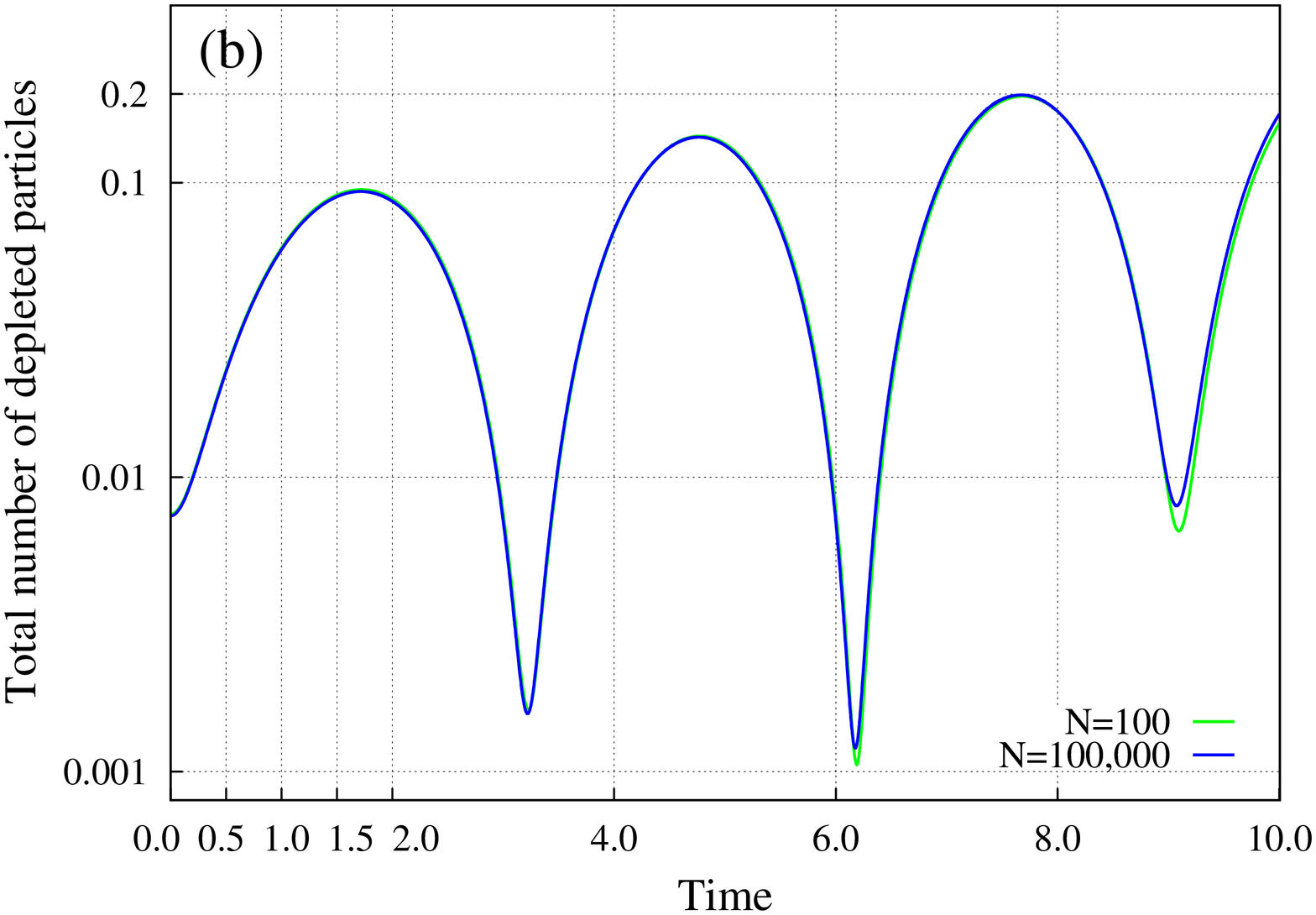}
\hglue -1.0 truecm
\includegraphics[width=0.365\columnwidth,angle=-90]{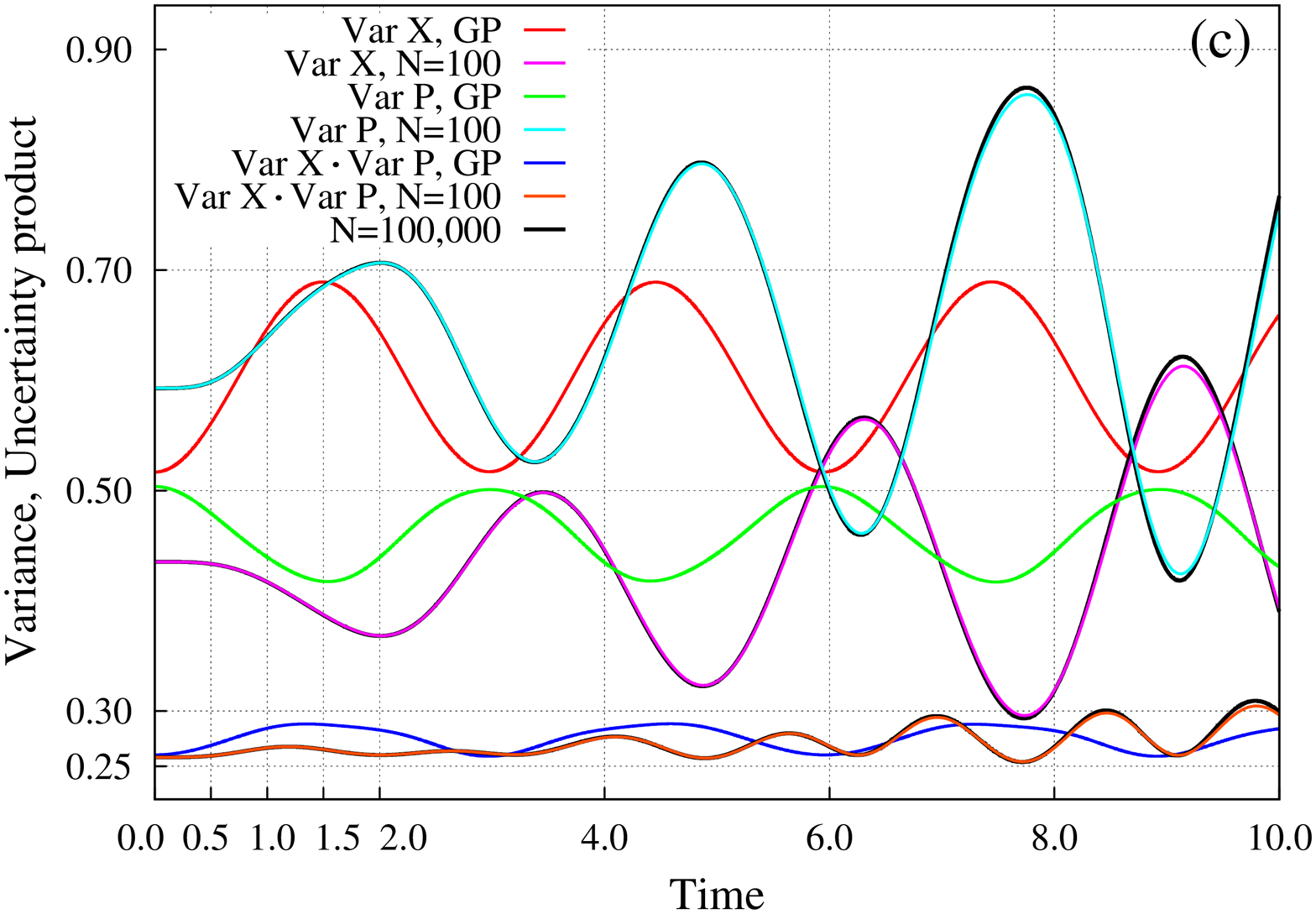}
\vglue 0.75 truecm
\caption{(Color online) 
Breathing dynamics following an interaction quench.
Shown and compared are many-body results for $N=100$ (using $M=4$ time-adaptive orbitals) 
and $N=100,000$ (using $M=2$ time-adaptive orbitals) bosons, 
and the mean-field results (equivalent to $M=1$ time-adaptive orbitals). 
The one-body Hamiltonian is $-\frac{1}{2}\frac{\partial^2}{\partial x^2} + \frac{x^4}{4}$
and the inter-particle interaction is harmonic, $\lambda_0\hat W(x_1-x_2)=-\lambda_0(x_1-x_2)^2, \lambda_0>0$.
(a) Snapshots of the density per particle, $\frac{\rho(x;t)}{N}$, as a function of time
following a sudden increase of the interaction parameter
from $\Lambda=\lambda_0(N-1)=0.19$ to $0.38$ at $t=0$.
The inset shows the density per particle at instances, from top to bottom, $t=0.0$, $0.5$, $1.0$, and $1.5$ for $N=100$ bosons.
(b) Total number of depleted particles outside the condensed mode 
as a function of time.
Note the values on the $y$ axis.
(c) Time-dependent many-particle position variance per particle,
$\frac{1}{N}\Delta^2_{\hat X}(t)$,
momentum variance,
$\frac{1}{N}\Delta^2_{\hat P}(t)$,
and uncertainty product,
$\frac{1}{N}\Delta^2_{\hat X}(t) \frac{1}{N}\Delta^2_{\hat P}(t) \equiv \Delta^2_{\hat X_{CM}}(t) \Delta^2_{\hat P_{CM}}(t)$.
Note the opposite behavior of the variance at short times
when computed at the many-body and mean-field level. 
See the text for more details.
The quantities shown are dimensionless.}
\label{f1}
\end{figure}

We consider $N=100$ and separately $N=100,000$ trapped bosons in one spatial dimension.
The one-body Hamiltonian is $-\frac{1}{2}\frac{\partial^2}{\partial x^2} + \frac{x^4}{4}$
and the inter-particle interaction is harmonic, $\lambda_0\hat W(x_1-x_2)=-\lambda_0(x_1-x_2)^2, \lambda_0>0$.
The system is prepared in the ground state of the trap for the interaction parameter
$\Lambda=\lambda_0(N-1)=0.19$.
At $t=0$ the interaction parameter is suddenly quenched to $\Lambda=0.38$ and we 
inquire what the out-of-equilibrium dynamics of the system would be like.
Fig.~\ref{f1} collects the results.

Fig.~\ref{f1}a depicts snapshots of the density per particle, $\frac{\rho(x;t)}{N}$, as a function of time.
The Gross-Pitaevskii and many-body results are seen to match very well.
The density is seen to perform breathing dynamics \cite{Bonz,Peter_2013,MCTDHB_3D_dyn}.
Since the interaction is repulsive and at $t=0$ quenched up,
the density first expands at short times.
In Fig.~\ref{f1}b the total number of depleted particles outside the condensed mode [$\alpha_1(x;t)$ natural orbital] 
are shown as a function of time. 
The systems are essentially fully condensed with only a fraction of a particle depleted.
In Fig.~\ref{f1}c the time-dependent many-particle position variance per particle,
$\frac{1}{N}\Delta^2_{\hat X}(t)$,
momentum variance,
$\frac{1}{N}\Delta^2_{\hat P}(t)$,
and uncertainty product,
$\frac{1}{N}\Delta^2_{\hat X}(t) \frac{1}{N}\Delta^2_{\hat P}(t) \equiv \Delta^2_{\hat X_{CM}}(t) \Delta^2_{\hat P_{CM}}(t)$
are shown.
Note the opposite behavior of the variance at short times
when computed at the many-body and Gross-Pitaevskii level.
Despite the expansion of the cloud (at short times),
the time-dependent position variance increases and momentum variance decreases,
implying that $\Delta_{\hat x, MB}^2(t)$ and $\Delta_{\hat p, MB}^2(t)$
are opposite in sign with respect to and dominate $\Delta_{\hat x, density}^2(t)$ and $\Delta_{\hat p, density}^2(t)$.
This is an interesting time-dependent many-body effect.   

\subsection{Pathway from condensation to fragmentation}\label{Ex2}

\begin{figure}[!]
\hglue -1.0 truecm
\includegraphics[width=0.345\columnwidth,angle=-90]{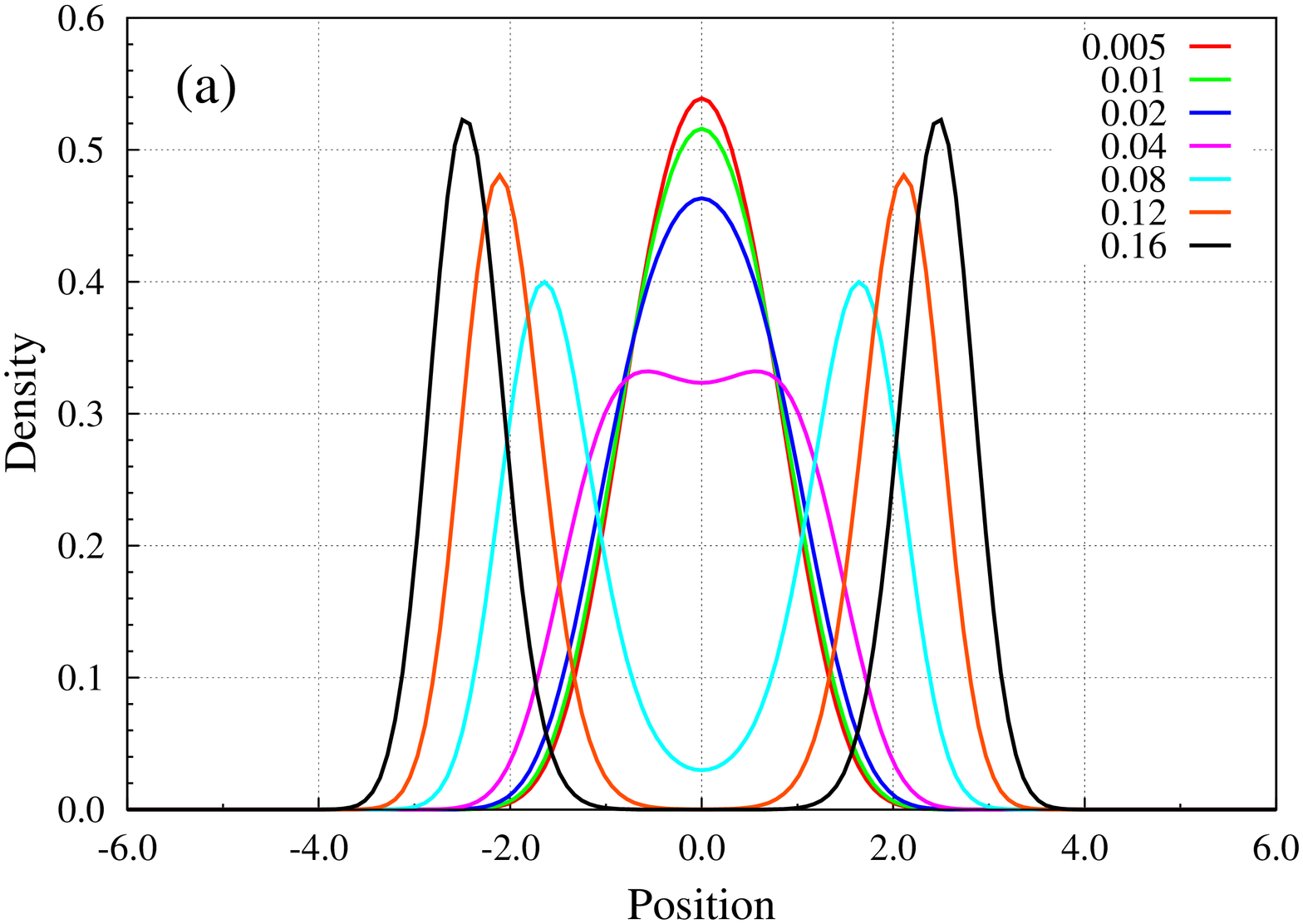}
\includegraphics[width=0.345\columnwidth,angle=-90]{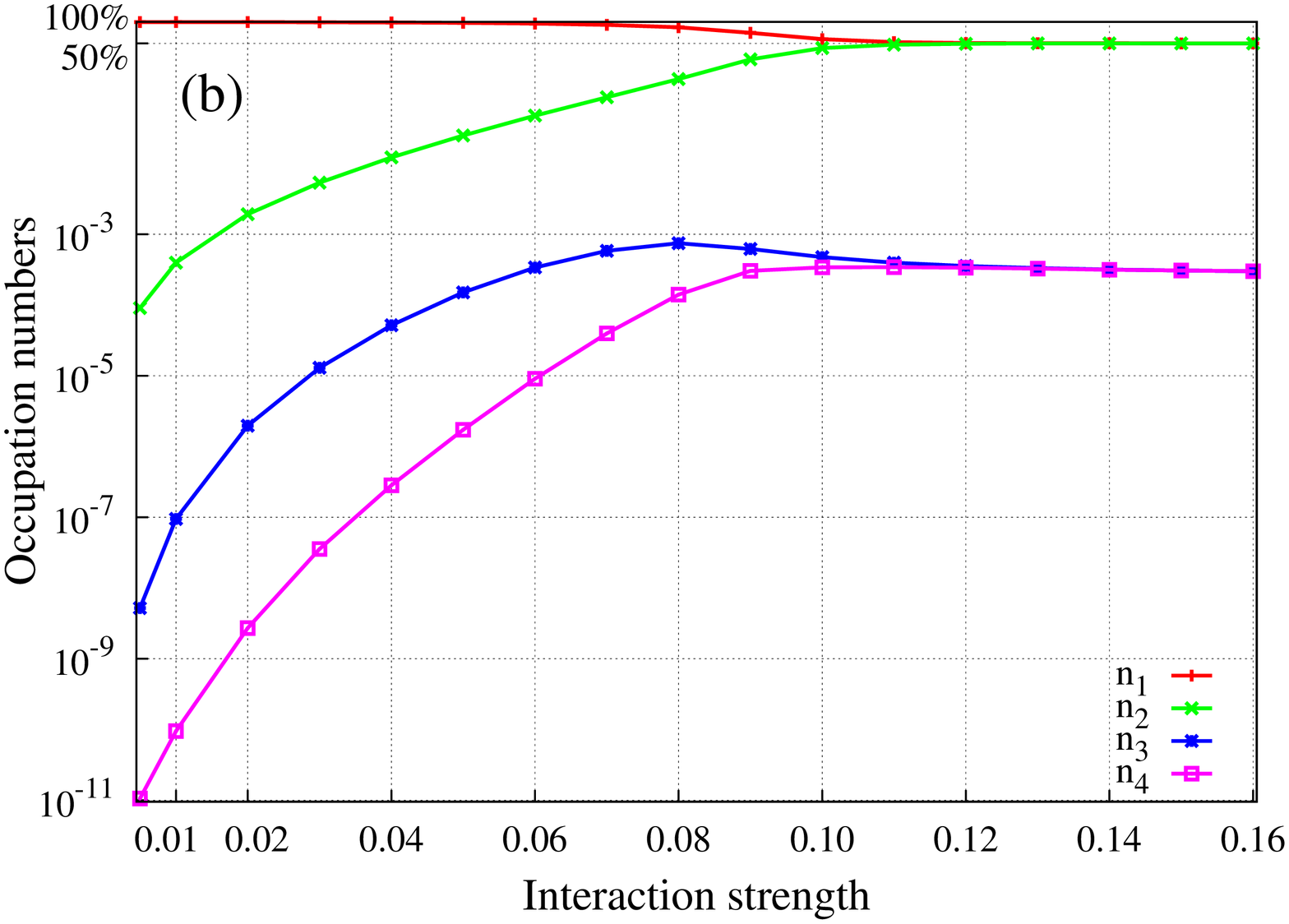}
\hglue -1.0 truecm
\includegraphics[width=0.365\columnwidth,angle=-90]{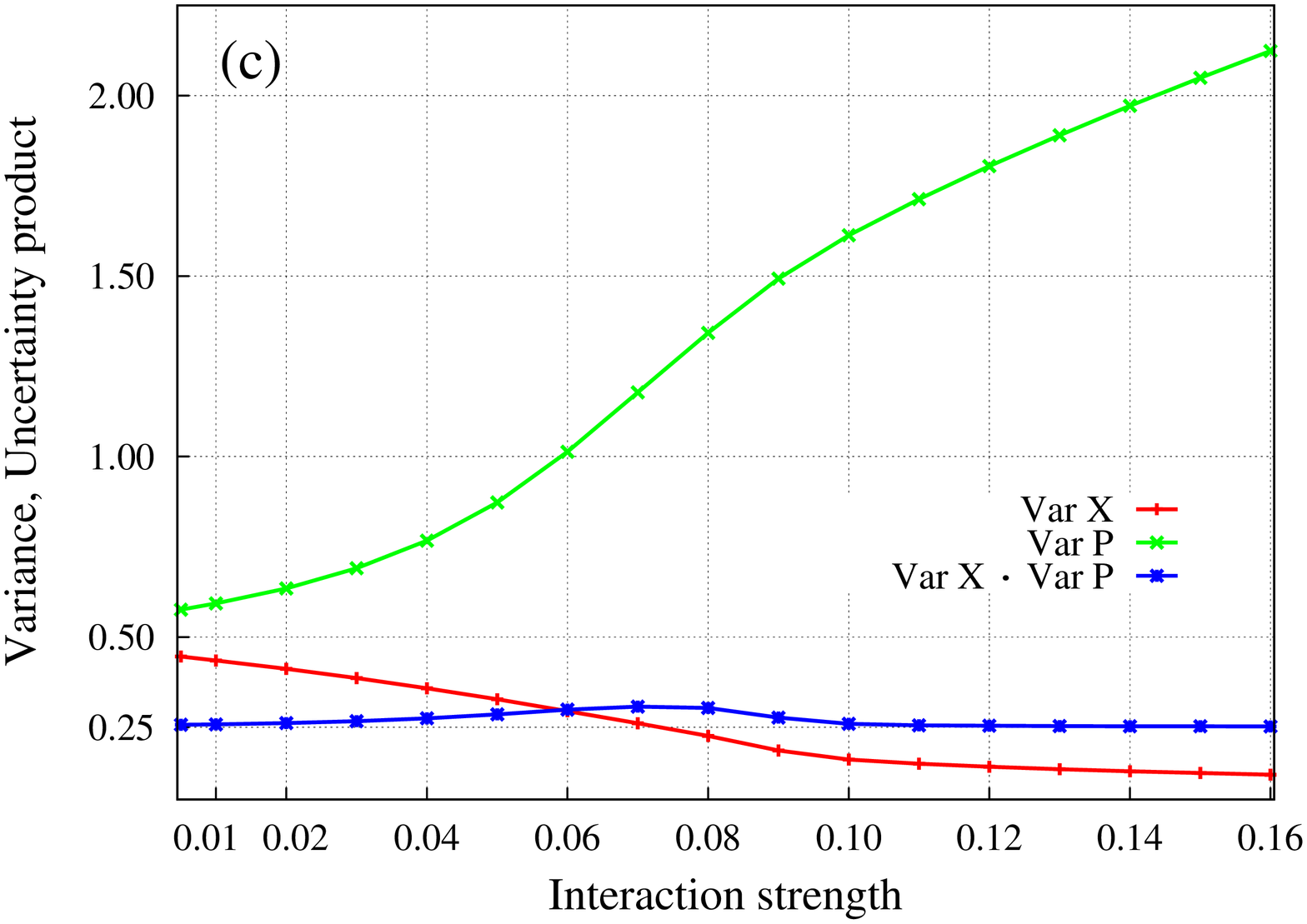}
\vglue 0.75 truecm
\caption{(Color online)
Pathway from condensation to fragmentation of the ground state.
Shown are many-body results for $N=20$ bosons (using $M=4$ time-adaptive orbitals). 
The one-body Hamiltonian is $-\frac{1}{2}\frac{\partial^2}{\partial x^2} + \frac{x^4}{4}$
and the inter-particle interaction is harmonic, $\lambda_0\hat W(x_1-x_2)=-\lambda_0(x_1-x_2)^2, \lambda_0>0$.
(a) Snapshots of the density per particle, $\frac{\rho(x;t)}{N}$, as a function of the interaction
strength $\lambda_0=0.005, \ldots, 0.16$.
As the interaction is increased the density broadens and splits into two parts.
(b) The four largest occupation numbers as a function of the interaction
strength $\lambda_0$ (points' symbols are data; smooth curves are to guide the eye).
The ground state essentially evolves from 100\% condensed to 50\%--50\% fragmented.
(c) Many-particle position variance per particle,
$\frac{1}{N}\Delta^2_{\hat X}$,
momentum variance,
$\frac{1}{N}\Delta^2_{\hat P}$,
and uncertainty product,
$\frac{1}{N}\Delta^2_{\hat X} \frac{1}{N}\Delta^2_{\hat P} \equiv \Delta^2_{\hat X_{CM}} \Delta^2_{\hat P_{CM}}$.
Note that, although the cloud's density broadens while the system's fragments,
the position variance actually decreases whereas the momentum variance increases.
See the text for more details.
The quantities shown are dimensionless.}
\label{f2}
\end{figure}

We next study the pathway from condensation to fragmentation of the ground state \cite{Sipe,ALN,Pathway,Pathway_Pe,MCHB}.
Fragmentation of BECs has drawn much attention, see, e.g., \cite{frg1,frg2,frg3,frg4,frg5,frg6,frg7,frg8}.
In particular for structureless bosons with a long-range interaction in a single-trap,
the ground state has been shown to become fragmentation when increasing the inter-particle repulsion \cite{MCTDHB_3D_dyn,Uwe_PRL1,MCTDHB_3D_stat,Uwe_PRL2,Uwe_PRL3}.
Fig.~\ref{f2} depicts the results for $N=20$ bosons.
The one-body Hamiltonian is again $-\frac{1}{2}\frac{\partial^2}{\partial x^2} + \frac{x^4}{4}$
and the inter-particle interaction is harmonic, $\lambda_0\hat W(x_1-x_2)=-\lambda_0(x_1-x_2)^2, \lambda_0>0$.
Fig.~\ref{f2}a shows snapshots of the density per particle, $\frac{\rho(x;t)}{N}$, as a function of the interaction strength $\lambda_0$. 
As the interaction is increased the density broadens and splits into two parts;
Side by side, the ground state fragments \cite{MCTDHB_3D_dyn,Uwe_PRL1,MCTDHB_3D_stat,Uwe_PRL2,Uwe_PRL3},
see Fig.~\ref{f2}b.
The ground state essentially evolves from 100\% condensed to 50\%--50\% fragmented.
Finally,
Fig.~\ref{f2}c displays the many-particle position variance per particle,
$\frac{1}{N}\Delta^2_{\hat X}$,
momentum variance,
$\frac{1}{N}\Delta^2_{\hat P}$,
and uncertainty product,
$\frac{1}{N}\Delta^2_{\hat X} \frac{1}{N}\Delta^2_{\hat P} \equiv \Delta^2_{\hat X_{CM}} \Delta^2_{\hat P_{CM}}$
of the ground state as a function of the interaction strength.
We find that, although the cloud's density broadens while the system's fragments,
the position variance decreases and the momentum variance increases,
unlike from what one would expect by just examining the density.
When the variances are combined,
the uncertainty product exhibits a maximum along the pathway from condensation to fragmentation.
This is an interesting static many-body effect.   

\section{Summary and Conclusions}\label{Conclusions}

We have studied in the present work
the variance and uncertainty product of
the position and momentum many-particle operators
of structureless bosons
interacting by a long-range inter-particle interaction 
and trapped in an anharmonic single-well potential.
There is no analytical solution to the many-particle Schr\"odinger equation of this system,
not even to the variance and uncertainty product themselves,
which makes a numerical solution of the out-of-equilibrium dynamics a must.
In the out-of-equilibrium interaction-quench scenario,
we have found that, despite the system being fully condensed, 
already when a fraction of a particle is depleted differences
with respect to the mean-field quantities arise.
In the static pathway from condensation to fragmentation scenario,
we found out that, although the cloud's density broadens while the system's fragments,
the position variance decreases, the momentum variance increases,
and the uncertainty product exhibits a maximum.
Both scenarios suggest a richness of effects emanating from the many-body
term of the variance and uncertainty product in interacting trapped many-boson systems.
Such many-body effects need not coincide with the information that can
be extracted based on the system's density alone.

\section*{Acknowledgements}

This research was supported by the Israel Science Foundation
(Grant No. 600/15). Partial financial support by the
Deutsche Forschungsgemeinschaft (DFG) is acknowledged.
We thank Lorenz Cederbaum for discussions. Computation
time on the Cray XC40 
system Hazelhen at the High Performance Computing Center
Stuttgart (HLRS) is gratefully acknowledged.

\appendix

\section*{Appendix: Details and convergence of the numerical computations}\label{APP_Num}

\begin{figure}[!]
\hglue -1.0 truecm
\includegraphics[width=0.345\columnwidth,angle=-90]{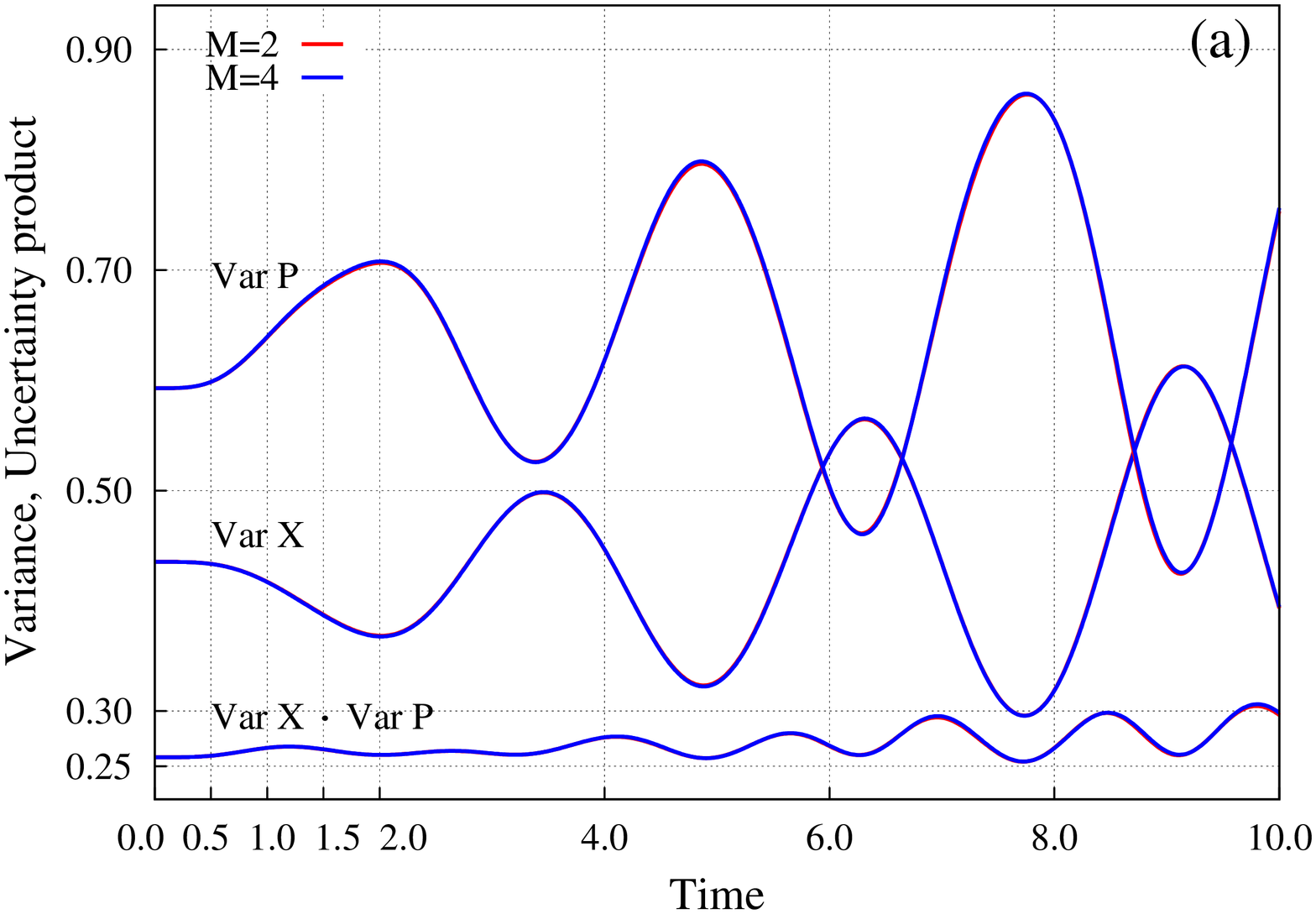}
\includegraphics[width=0.345\columnwidth,angle=-90]{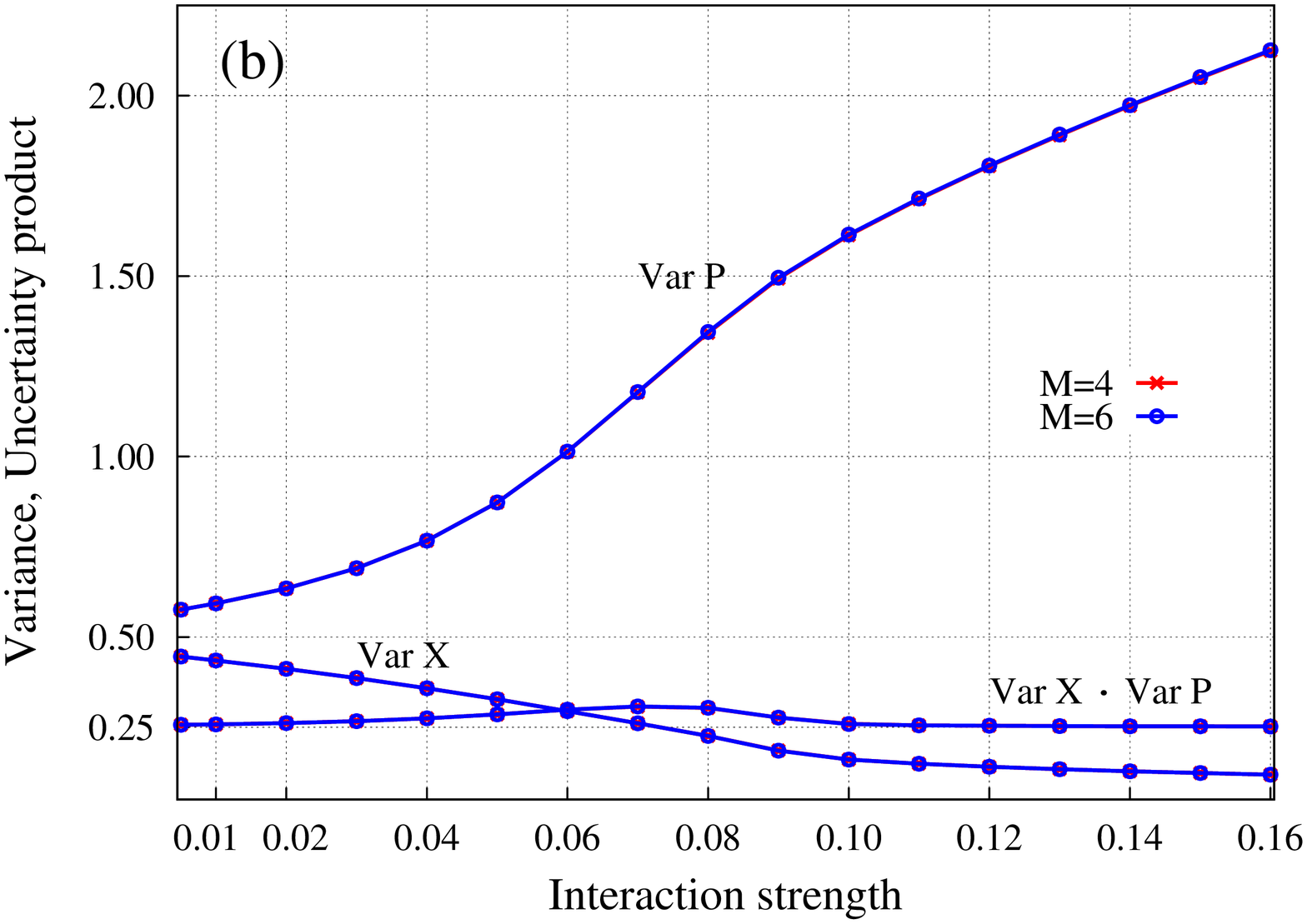}
\vglue 0.75 truecm
\caption{(Color online) 
Convergence of the many-particle position variance per particle,
$\frac{1}{N}\Delta^2_{\hat X}(t)$,
momentum variance,
$\frac{1}{N}\Delta^2_{\hat P}(t)$,
and uncertainty product,
$\frac{1}{N}\Delta^2_{\hat X}(t) \frac{1}{N}\Delta^2_{\hat P}(t) \equiv \Delta^2_{\hat X_{CM}}(t) \Delta^2_{\hat P_{CM}}(t)$
with the number of time-adaptive orbitals $M$ used in the MCTDHB computations for the systems consisting of $N=100$
and $N=20$ bosons discussed in the main text.
(a) The out-of-equilibrium interaction-quench breathing dynamics in Sec.~\ref{Ex1}
(propagation of the MCTDHB equation of motion in real time).
It is found that the results with $M=2$ and $M=4$ orbitals lie atop each other. 
(b) The pathway from condensation to fragmentation of the ground state in Sec.~\ref{Ex2} 
(propagation of the MCTDHB equations of motion in imaginary time). 
It is found that the results with $M=4$ and $M=6$ orbitals lie atop each other. 
The quantities shown are dimensionless.}
\label{f3}
\end{figure}

The multiconfigurational time-dependent Hartree for bosons 
(MCTDHB) method \cite{MCTDHB1,MCTDHB2,book_MCTDH,Kaspar_The,book_nick,Axel_The} 
is used in the present work to compute the
time-dependent and ground-state properties
of trapped bosons interacting by a long-range inter-particle interaction.
To obtain the ground state
we propagate the MCTDHB equations of motion in imaginary time \cite{Benchmarks,MCHB}. 
For the computations the many-body Hamiltonian is represented 
by $256$ exponential discrete-variable-representation grid points
(using a Fast-Fourier Transform routine) in a box of size $[-10,10)$.
We use the numerical implementation in the software packages \cite{MCTDHB_LAB,package}.
Convergence of the variance and uncertainty product with increasing number $M$ 
of time-adaptive orbitals is demonstrated for $N=100$ bosons in Fig.~\ref{f3}a for the out-of-equilibrium
dynamics \cite{TD_Variance} and for $N=20$ bosons in Fig.~\ref{f3}b for the ground state \cite{Variance},
also see in this context \cite{Brand_Cos}.
It is found that, respectively, 
the results with $M=2$ and $M=4$ orbitals are converged.

\end{document}